\documentclass{tran-l}
\usepackage{amsmath, amssymb}
\usepackage{graphicx}
\usepackage{color}
\usepackage{textcomp}
\usepackage{hyperref}

\usepackage{graphics}

 \newtheorem{thm}{Theorem}[subsection]

 \newtheorem{prop}[thm]{Proposition}
 \theoremstyle{definition}
 
 \theoremstyle{remark}
 \newtheorem{rem}[thm]{Remark}
 \numberwithin{equation}{section}

\newcommand{\bm}{\bibitem}

\newcommand{\be}{\begin{equation}}
\newcommand{\ee}{\end{equation}}

\newcommand{\bea}{\begin{eqnarray}}
\newcommand{\bes}{\begin{subequations}}
\newcommand{\ees}{\end{subequations}}
\newcommand{\bgt}{\begin{gather}}
\newcommand{\egt}{\begin{gather}}
\newcommand{\eea}{\end{eqnarray}}
\newcommand{\beaa}{\begin{eqnarray*}}
\newcommand{\eeaa}{\end{eqnarray*}}

\newcommand{\HH}{{\mathbb H}}

\newcommand{\cal}{\mathcal}
\newcommand{\BB}{{\mathbb B}}
\newcommand{\QQ}{{\mathbb Q}}

\begin{document}

\title[Unifying the BGM and SABR Models]
{Unifying the BGM and SABR Models: \\ A short Ride in Hyperbolic
Geometry}

\author{Pierre Henry-Labord\`ere}
\address{Soci\'et\'e G\'en\'erale, Equity Derivatives Research, Paris, France.}
\email{pierre.henry-labordere@sgcib.com}


\maketitle

\begin{abstract}
In this short note, using our geometric method introduced in a
previous paper \cite{phl} and initiated by \cite{ave}, we derive
an asymptotic swaption implied volatility at the first-order for a
general stochastic volatility Libor Market Model. This formula is
useful to quickly calibrate  a  model to a full swaption matrix.
We apply this formula to a specific model where the forward rates
are assumed to follow a multi-dimensional  CEV process correlated
to a SABR process. For a caplet, this model degenerates to the
classical SABR model and our asymptotic swaption implied
volatility reduces naturally  to the Hagan-al formula \cite{sab}.
The geometry underlying this model is the hyperbolic manifold
$\HH^{n+1}$ with $n$ the number of Libor forward rates.
\end{abstract}


\section{Introduction}

The BGM model \cite{bgm,jam} has recently been  the focus of much
attention as it gives a theoretical justification for pricing
caps-floors using the classical Black-Scholes formula. The basic
(physical) random variables are given by the Libor forward rates
which are assumed to follow a correlated log-normal process.
\noindent As the forward swap rate model implied by the BGM is
quite complicated (the swap forward  rate is not log-normally
distributed), the calibration to the swaption matrix is difficult.
Asymptotic swaption implied volatility (at the zero-order) were
initially derived by Rebonato \cite{reb} and Hull-White \cite{hul}
for the (log-normal) BGM model. Such formula has been obtained by
assuming that the ratio of a forward Libor rate over the swap rate
and the derivative of the swap rate according to a forward Libor
rate are almost constant (and therefore equal to their values at
the spot).

\noindent Despite its great success, the BGM model presents the
same drawbacks as the classical Black-Scholes theory: As the
forward rates follow a correlated log-normal process, the model is
not able to calibrate the full swaption matrix in/out-the money
(in particular the caplets) and give a good dynamics to the Libor
rates. The incorporation of a swaption smile can be obtained by
introducing more elaborated models which should be flexible enough
to calibrate caplets and a grid of swaption volatilities (not
necessary at the money) across all swaption expiries and
underlying swap maturities. One property that these models must
still share is their ability to quickly calibrate  the swaption
matrix without using complicated numerical routines such as
Monte-Carlo simulation which are usually noisy and time-consuming.
In this context, Andersen-Andreasen introduced the CEV Libor
Market Model (LMM) \cite{and} which assumes that each forward rate
follows a CEV process, and showed how to obtain asymptotic
swaption smile. Their method is still based on the Rebonato
"freezing" argument which is not completely mathematically
justified. Recently, for this specific model, Kawai found a better
asymptotic formula using the Wiener chaos expansion \cite{kaw}.
Although giving more flexibility than the BGM model, the CEV LMM
model is still not able to calibrate the swaption matrix for
in/out strike and in this context, we are naturally led to use
stochastic volatility LMM. The literature on this subject is not
particularly large. Andersen-al introduced a LMM where the Libors
follow a multi-dimensional correlated CEV process  coupled (but
uncorrelated) to a Heston model \cite{and2,and3} and recently V.
Piterbarg modifies this model to incorporate term structure
\cite{pit}. Using an averaging principle, V. Piterbarg derives an
asymptotic volatility. Note that as these models are uncorrelated
to the stochastic volatility, the swaption fair value is simply
given by the fair price in the case of a local volatility model
conditional to the stochastic volatility process as explained by
the Hull-White decomposition \cite{hw2}. An asymptotic expression
can then be generated by approximating the moments of the
volatility process\cite{and2}.

\noindent For pricing exotic options (such as bermudan swaptions
for example), it is simpler or more natural to model directly the
forward swap rate with a stochastic volatility process. For
example, the SABR model \cite{sab} was introduced to fulfill his
goal. An asymptotic swaption smile formula (at the first-order)
was derived for this specific model and help to calibrate quickly
the model to liquid market data. In this context, it is  natural
to try to reconcilate/unify the two benchmark models, the BGM and
SABR models. We therefore introduce a LMM where the forward rates
follow a multi-dimensional  CEV process (with one beta for each
forward) correlated to a SABR model. As it is the case for the
SABR model, we impose that the libors are correlated to the unique
volatility and it is therefore not possible to follow the
Andersen-al \cite{and3} method (i.e. the Hull-White decomposition)
to derive an asymptotic swaption smile.

\noindent In this paper, we  pursue our previous work on the
application of the heat kernel expansion on a Riemannian manifold
endowed with an Abelian connection \cite{phl} to derive an
asymptotic smile formula for a swaption. The plan of this paper is
as follows: In the first part, we will recall some definitions and
present a list of  recent Libor Market Models. In the second part,
we apply this heat kernel expansion to derive an asymptotic
swaption smile formula at the first-order valid \underline{for
any} LLM. In the third part, we present our stochastic LMM and
apply this general formula. We will prove that the geometry
underlying this model is the hyperbolic manifold $\HH^{n+1}$. Some
important properties of this space are then presented.
Furthermore, we show that the "freezing" argument is no longer
valid when we try to price a swaption in/out the money: The libors
should in fact be frozen to the saddle-point (constrained on a
particular hyperplane) which minimizes the geodesic distance on
$\HH^{n+1}$.

\section{Libor Market Model}

A swaption gives the right, but not the obligation, to enter into
an interest rate swap at a pre-determined rate on an agreed future
date \cite{bri}. The maturity date for the swaption  is noted
$T_\alpha$ and $T_\beta$ is the expiry for the forward swap rate
$s_{\alpha \beta}$ given by \bea s_{\alpha
\beta}(t)={1-\prod_{j=\alpha+1}^\beta {1 \over (1+\tau F_j(t))}
\over \sum_{i=\alpha+1}^\beta \tau \prod_{j=\alpha+1}^i {1 \over
(1+\tau F_j(t))}} \eea

\noindent with $\tau$ the tenor and $F_k(t)\equiv
F(t,T_{k-1},T_k)$ is the forward rate resetting at $T_{k-1}$.

\noindent As the product of the bond $P(t,T_{k})$ with the forward
rates $F_k(t)$ is a difference of two bonds with maturity
$T_{k-1}$ and $T_k$, ${1 \over \tau}(P(t,T_{k-1})-P(t,T_{k})$, and
therefore a traded asset, $F_k$ is a (local) martingale under
$\QQ^k$, the (forward) measure associated with the num\'eraire
$P(t,T_k)$. Therefore, we assume  the following driftless dynamics
\bes \bea
 dF_k(t)&=&\sigma_k(t) \Phi_k(a,F_k) dW_k \; , \; \forall t \leq
T_{k-1} \; , \; k=1,\cdots,n \label{sde1} \\
dW_kdW_l&=&\rho_{kl}(t) dt \label{sde2} \eea \ees \noindent with
the initial conditions $a(t=0)=\alpha$ and $F_k(t=0)=F_k^0$.

\begin{table}[h]
\begin{center}
\begin{tabular}{|c|c|}
\hline Libor market model & SDE \\
\hline BGM &  $dF_k=\sigma_k(t) F_k dW_k$ \\
\hline CEV & $dF_k=\sigma_k(t) F_k^\beta dW_k$ \\
\hline Limited CEV & $dF_k=\sigma_k(t) F_k min(F_k^{\beta-1},\epsilon^{\beta-1})  dW_k$  \\
                    & with $\epsilon$ a small positive number \\
\hline Shifted log-normal & $dX_k=\sigma(t) X_k dW_k$  \\
                    & with $F_k=X_k+\alpha$ \\
\hline FL-SV &  $dF_k=\sigma_k(t) ( \beta F_k +(1-\beta)F_k^0 )  \sqrt{v} dW_k$  \\
                    &  $dv=\lambda(v-\bar{\lambda})dt+\nu \sqrt{v} dZ$ ; $dW_k dZ=0$ \\

\hline FL-TSS &  $dF_k=\sigma_k(t) ( \beta_k(t) F_k +(1-\beta_k(t))F_k^0 )  \sqrt{v} dW_k$  \\
                    &  $dv=\lambda(v-\bar{\lambda})dt+\nu \sqrt{v} dZ$ ; $dW_k dZ=0$ \\

\hline
\end{tabular}
\caption{Examples of stochastic (or local) volatility Libor
models.}
\end{center}
\end{table}

\noindent In order to achieve some flexibility, we assume that the
(normal) local volatility $\Phi_k(a,F_k)$ depends on a hidden
Markov process $a$ (to be specified later) representing a
stochastic volatility. We therefore assume that all the forward
rates are coupled with the same stochastic volatility $a$. (Table
1) presents a list of the different functional forms for $\phi_k$
used in the literature. The BGM, (limited) CEV and shifted
log-normal models correspond to local volatility models ($a=1)$
and the others to  stochastic volatility models with a unique
stochastic volatility $a$ driven by a Heston process. Note that
the stochastic differential equation for the libors $L_k$ has been
written in the forward measure $\QQ^k$ and the stochastic equation
for $a$ remains the same in the forward or forward swap rate
measures as $a$ is assumed to be uncorrelated with the Libor
rates. This will not be the case  in our LLM.

\section{Asymptotic Swaption Smile}

\noindent The forward swap rate satisfies the following driftless
dynamics in the forward-swap measure $\QQ^{\alpha \beta}$
(associated to the num\'eraire $C_{\alpha
\beta}(t)=\sum_{i=\alpha+1}^\beta \tau_i P(t,T_i)$) \bea
ds_{\alpha \beta}=\sum_{k=\alpha+1}^\beta {\partial s_{\alpha
\beta} \over
\partial F_k} \sigma_k(t) \phi_k(a,F_k) dZ_k \eea

\noindent with ${\partial s_{\alpha \beta} \over \partial F_j}= {
s_{\alpha \beta} \tau_j \over (1+\tau_j F_j)}( { P(t,T_\alpha)
\over P(t,T_\beta)-P(t,T_\alpha)}+{ \sum_{k=j}^{\alpha-1} \tau_k
P(t,T_{k+1}) \over C_{\alpha \beta}})$

\noindent The local volatility associated to the forward swap rate
($ds_{\alpha \beta}=\sigma_{loc}^{\alpha \beta}(s_{\alpha
\beta},t) dW_t$) is then by definition
\bes \bea
(\sigma_{loc}^{\alpha \beta})^2(s,t)&\equiv& {\mathbb E}^{\alpha
\beta}[ \sum_{i,j=\alpha+1}^\beta \rho_{ij}(t) \sigma_i(t)
\sigma_j(t)\phi_i(a,F_i) \phi_j(a,F_j)
 {\partial s_{\alpha \beta} \over
\partial F_i} {\partial s_{\alpha \beta} \over \partial
F_j}|s_{\alpha \beta}=s]\\
 &=&\sum_{i,j=\alpha+1}^\beta \rho_{ij}(t) \sigma_i(t)
\sigma_j(t) {\int_{\BB} \phi_i(a,F_i) \phi_j(a,F_j)
 {\partial s_{\alpha \beta} \over
\partial F_i} {\partial s_{\alpha \beta} \over \partial F_j} p
da \prod_i dF_i  \over \int_{\BB} p da \prod_i dF_i} \label{lv}
\eea \ees \noindent with the submanifold $\BB=\{\{F_i\}_i|
s_{\alpha \beta}=s\}$ and $p\equiv p(a,{F}_i,t|\alpha,{F}_i^0)$
the conditional probability satisfying the (backward) Kolmogorov
equation associated to the SDE for the Libors and the volatility
$a$ in the forward swap measure $\QQ^{\alpha \beta}$. An
asymptotic expression in the short time limit for the local
volatility $\sigma_{\alpha \beta}(s,t)$ can be found in two steps:
find an asymptotic expansion for the conditional probability $p$
(in  $\QQ^{\alpha \beta}$) and do the integration over ${\mathbb
B}$.

\subsection{First step: Heat Kernel expansion}

As explained previously, the first step can be achieved using the
heat kernel expansion. In that purpose, the Kolmogorov equation is
rewritten as the heat kernel equation on a ($n+1$)-dimensional
Riemannian manifold ${\mathcal M}^{n+1}$ endowed with an Abelian
connection as explained in \cite{phl,phl2,phl3}. Let's assume that
our multi-dimensional stochastic equations (in  $\QQ^{\alpha
\beta}$) are written as \bea
dx^\mu=b^\mu(x,t)dt+\sigma^\mu(x,t)dW^\mu \eea \noindent with
$dW^\mu dW^\nu=\rho_{\mu \nu}(t)dt$ (note that the indices
$1,\cdots,n$ (resp. $n+1$) correspond(s) to the forward $F^i$
(resp. a)). Then, the metric $g_{\mu \nu}$ depends only on the
diffusion terms $\sigma_\mu$ and the connection ${\mathcal A}_\mu$
on the drift terms $b^\mu$ as well \bes \bea g_{\mu \nu}(x,t)&=&2
{\rho^{\mu \nu}(t) \over \sigma_\mu(x,t) \sigma_\nu(x,t)} \; , \;
\mu,\nu=1 \cdots n+1\label{metric} \;, \; \rho^{\mu \nu} \equiv
[\rho^{-1}]_{\mu \nu}
 \\ {\mathcal A}^\mu(x,t)&=&{1 \over 2}(b^\mu(x,t)-\sum_{\nu=1}^{n+1} g^{-{1 \over
2}}\partial_\nu(g^{1/2}g^{\mu \nu}(x,t))) \;,\; \mu=1 \cdots n+1
 \label{aaa} \eea \ees
\noindent with $g(x,t) \equiv det[g_{\mu \nu}(x,t)]$. In terms of
these functions, the asymptotic solution to the Kolmogorov
equation in the short-time limit is given by ($x=(a,F_i)$,
$x^0=(\alpha, F_i^0)$)
 \bea
p(x,t|x^0)={\sqrt{g(x)} \over (4\pi t)^{n \over
2}}\sqrt{\Delta(x,x^0)}{\cal P}(x,x^0)e^{-{\sigma(x,x^0) \over
2t}}\sum_{n=1}^\infty a_n(x,x^0)t^n \;, \; t \rightarrow 0
\label{shk}\eea

\begin{itemize}

\item \noindent Here, $\sigma(x,x^0)$ is the Synge world function
equal to one half of the square of geodesic distance $d(x,x^0)$
between $x$ and $x^0$ for the metric $g_{\mu \nu}(x,t=0)$. This
distance is defined as the minimizer of \bea d(x,x^0)^2=min_{C}
\int_0^T g_{\mu \nu}(x,t=0) {dx^\mu \over dt} {dx^\nu \over dt} dt
\eea and $t$ parameterizes the curve ${\cal C}(x,x^0)$ joining
$x(t=0)\equiv x^0$ and $x(T)\equiv x$.

\item \noindent $\Delta(x,x^0)$ is the so-called Van Vleck-Morette
determinant \bea \Delta(x,x^0)=g(x,0)^{-{1 \over 2}}det (-
{\partial^2 \sigma(x,x^0) \over
\partial x
\partial x^0}  )g(x^0,0)^{-{1 \over 2}} \label{mor}\eea

\noindent with $g(x,0)=det[g_{\mu \nu}(x,0)]$

\item ${\cal P}(x,x^0)$ is the parallel transport of the Abelian
connection along the geodesic ${\cal C}(x,x^0)$ from the point
$x^0$ to $x$

\bea {\cal P}(x,x^0)=e^{\int_{C(x,x^0)} {\cal A}_\mu(x,t=0)
dx^\mu} \label{pargauge} \eea

\item \noindent The $a_i(x,x^0)$ ($a_0(x,x^0)=1$) are smooth
functions on $M$ and depend on geometric invariants such as the
scalar curvature $R$. More details can be found in \cite{phl}.

\end{itemize}

\subsection{Second step: Saddle-point method}

The integration over ${\mathbb B}$ is  obtained by using a
saddle-point method which consists in approximating at the first
order the integral $\int f(x) e^{ \epsilon \phi(x)} dx $ in the
limit $\epsilon$ small by \cite{erd}

\bea \int f(x) e^{ \epsilon \phi(x)} dx \sim_{\epsilon <<1} &&
f(x^*) e^{ \epsilon \phi(x^*)}(1+ {1 \over
\epsilon}(-{\partial_{\alpha \beta} f \over 2f} {A_{\alpha \beta}
} +( {\partial_\alpha f \over 2f}
\partial_{\beta \gamma \delta} \phi +{1 \over 8 }
\partial_{\alpha \beta \gamma \delta } \phi) A_{\alpha \beta}A_{\gamma \delta} \nonumber \\&&-{\partial_{\alpha \beta \gamma} \phi
\partial_{\delta \mu \nu} \phi \over 72} A_{\alpha \beta} A_{\gamma \delta} A_{\mu \nu})) \label{sad}\eea

\noindent with $A^{\alpha \beta}=[\partial_{\alpha \beta}
\phi]^{-1}$, $dx \equiv \prod_{i=1}^n dx_i$ and $x^*$ the
saddle-point (which minimizes $\phi(x)$). This expression can be
obtained by developing $\phi(x)$ and $f(x)$ in series around
$x^*$. The quadratic part in $\phi(x)$ leads to a Gaussian
integration over $x$ which can be performed.

\subsubsection{Saddle-point}

As the conditional probability at the zero-order is proportional
to $e^{-{d(x,x^0)^2 \over 4t}}$, the saddle-point corresponds to
the point $x=(a,{F}_{i=1,\cdots,n})$ on the submanifold $s_{\alpha
\beta}=s$ which minimizes the geodesic distance $d(x,x^0)$
\cite{ber,ave} \bea (a^*,\{F_i^*\})\equiv (a,\{F_i \}) \;
{\mathrm such \; as} \; min_{a, \{F_i\}| s_{\alpha \beta}=s }[
d(x,x^0)^2] \label{local0} \eea

\noindent Introducing a Lagrange multiplier, $\lambda$, this is
equivalent to \bea (a^*,\{F_i^*\}) \equiv (a,\{F_i \}) \; {\mathrm
such \; as} \; min_{a, \{F_i\},\lambda}
[d^2(x,x^0)+\lambda(s_{\alpha \beta}(F)-s)] \label{local1} \eea

\subsection{Asymptotic local volatility}

\noindent Plugging our asymptotic expression for the conditional
probability (\ref{shk}) into (\ref{sad}), we finally obtain the
local volatility at the first-order \bea (\sigma_{loc}^{\alpha
\beta})^2(s,t)=\sum_{i,j=1}^n \rho_{ij}(t) \sigma_i(t) \sigma_j(t)
f_{ij}(F^*,a^*) (1+2t \sum_{\mu,\nu=1}^{n+1}A^{\mu \nu}  (
({\partial_{\mu \nu} f_{ij}(F^*,a^*) \over f_{ij}(F^*,a^*)} +
2{\partial_{\mu} f_{ij}(F^*,a^*) \over
f_{ij}(F^*,a^*)}{\partial_{\nu} \psi(F^*,a^*) \over
\psi(F^*,a^*)}) \nonumber
\\-\sum_{\gamma,\delta=1}^{n+1}  A^{\gamma
\delta} {\partial_{\mu} f_{ij}(F^*,a^*) \over f_{ij}(F^*,a^*)}
\partial_{\nu \gamma \delta} d^2 )) \label{localvol}
 \eea

\noindent with $f_{ij}(F,a)=a^2 C_i(F_i) C_j(F_j) {\partial
s_{\alpha \beta} \over \partial F_i} {\partial s_{\alpha \beta}
\over
\partial F_j}$, $\psi(F,a)=\sqrt{g \Delta}{\cal P}$ and $A^{\mu \nu}=[\partial_{\mu \nu}
d^2]^{-1}$.

\noindent Note that as opposed to other asymptotic methods
presented in the literature, this formula is \underline{exact} at
$t=0$. The zero-order formula (independent of the time $t$ for
$\sigma_i(t)$, $\rho_{ij}(t)$ constant) was derived for a general
multi-dimensional local volatility model by \cite{ave,ber}.
Moreover, in the expansion, we assumed that time is small but we
made no assumption that $F_k$ is close to the spot libor or that
the volatility of volatility is small.

\subsection{Asymptotic Smile}

The asymptotic smile can be derived in two steps from the
asymptotic local volatility: first, we have

\bea ds={{\sigma}_{loc}^{\alpha \beta}(s,t) \over
{\sigma}_{loc}^{\alpha \beta}(s_0,t)}{\sigma}_{loc}^{\alpha
\beta}(s_0,t)dW_t \eea

\noindent and doing a change of local time $t'=\int_0^t
{\sigma}_{loc}^{\alpha \beta}(s_0,u)^2du$, we now obtain the
associated local volatility model for the swap rate \bea
ds_{\alpha \beta}=\bar{\sigma}_{loc}^{\alpha \beta}(s,t)dW_t' \eea

\noindent with $\bar{\sigma}_{loc}^{\alpha
\beta}(s,t)={{\sigma}_{loc}^{\alpha \beta}(s,t) \over
{\sigma}_{loc}^{\alpha \beta}(s_0,t)}$.

\noindent Secondly, we know that there is a one-to-one
correspondence between this local volatility and the smile
\cite{phl}  given at the first-order by  ($C(f)  \equiv
\bar{\sigma}_{loc}^{\alpha \beta}(s,t=0)$) \bea \sigma^{\alpha
\beta}_{BS}(K,T_\alpha)=&&\sqrt{\int_0^{T_\alpha}
{\sigma}_{loc}^{\alpha \beta}(s_0,u)^2du \over  T  }{ ln({K \over
s_0})  \over \int_{s_0}^K {df' \over C(f')}}(1 +{C^2(f_{av})
\int_0^T {\sigma}_{loc}^{\alpha \beta}(s_0,u)^2du \over 24} ( 2{
C''(f) \over C(f_{av})}-({C'(f_{av}) \over C(f_{av})})^2+{1 \over
f_{av}^2} \nonumber
\\&&+ 12 {\partial_t \bar{\sigma}_{loc}^{\alpha \beta}(f_{av},t=0)
\over C^3(f_{av})} )|_{f_{av}={s_0+K \over 2}}) \label{smile}
 \eea

\section{SABR-LMM model}

\noindent We have seen that the asymptotic local and implied
volatilities can be
 computed if we know the geodesic distance and a parametrization
of geodesic curves on ${\mathcal M}^{n+1}$. This is the case  for
the hyperbolic space $\HH^n$ for all $n$. This manifold has a lot
of important properties. As such, it appears to be the perfect toy
model (usually its Lorentzian version AdS/dS) in a number of
domain: chaos, cosmology, string theory, .... In the first part,
we present our BGM-LLM-SABR model and show that the underlying
geometry  is $\HH^{n+1}$ (with $n$ the number of forward Libor
rates). Using this connection, we will find an asymptotic local
volatility and an asymptotic swaption implied volatility.

\subsection{Dynamics}

\noindent We introduce the model SABR-LMM, given by the following
SDE under the spot Libor measure $\QQ$ (associated to the
num\'eraire $B_d(t)=\prod_{j=1}^{\beta(t)-1}(1 +\tau_j F_j(T_j-1))
P(t,T_{\beta(t)-1})$ where $\beta(t)=m$ if $T_{m-2} <t<T_{m-1}$)
\bes \bea dF_k&=& a^2 B^k(F,t) dt +\sigma_k(t) a C_k(F_k) dZ_k  \label{libor} \\
da&=&\nu a dZ_{n+1} ;  \;, \; dZ_idZ_j=\rho_{ij}(t)dt \; \;
i,j=1,\cdots,n+1\eea \ees \noindent with \bea C_k(F_k)&=&\phi_k
F_k^{\beta_k}
\\ B^k(F,t)&=& \sum_{j=\beta(t)}^k {\tau_j \rho_{jk} \sigma_k(t) \sigma_j(t) C_k(F_i) C_i(F_k)\over
(1+\tau_j F_j)} \eea \noindent The functions $C_k(F_k)$ have been
scaled by $\phi_k$ and therefore we can impose that
$\sigma_k(0)=0$. The stochastic equation for $a$ was written in
the spot Libor measure in order to get a SDE of a specific
underlying swap $s_{\alpha \beta}$ or a forward bond. Under the
forward swap measure $\QQ^{\alpha \beta}$, we have \bes
\bea dF_k&=& a^2 b^k(F,t) dt +\sigma_k(t) a C_k(F_k) dZ_k  \label{libor} \\
da&=&-\nu^2 a^2 b^a(F,t) + \nu a dZ_{n+1} ;  \;, \;
dZ_idZ_j=\rho_{ij}(t)dt \; \; i,j=1,\cdots,n+1\eea \ees
\noindent
with \bea b^k(F,t)&=& \sum_{j=\alpha+1}^\beta (2 1_{j\leq
k}-1)\tau_j {P(t,T_j) \over C_{\alpha
\beta}(t)}\sum_{i=min(k+1,j+1)}^{max(k,j)}{ \tau_i \rho_{ki}
\sigma_i(t) \sigma_k(t) C_i(F_i) C_k(F_k)\over (1+\tau_i F_i)} \\
C_{\alpha \beta}(t)&=&\sum_{i=\alpha+1}^\beta \tau_i P(t,T_i) \\
b^a(F,t)&=& \sum_{i=\alpha+1}^\beta \tau_i \omega_i(t)
\sum_{k=1}^i {\tau_k C_k(F_k) \rho_{ka} \sigma_k(t) \over 1
+\tau_k F_k(t)} \eea

\noindent and with $\omega_i(t)={\prod_{k=1}^i {1 \over (1+\tau_k
F_k)} \over \sum_ {j=\alpha+1}^\beta \prod_{k=1} ^j {1 \over
(1+\tau_k F_k)}}$

\noindent Note that the forward-rate dynamics under the forward
measure $\QQ^{k}$ is much simpler and  given by the following
stochastic differential equations (SDE) \bea dF_k(t)=\sigma_k(t) a
C_k(F_k) dW_k   \;, \;  dW_k dW_p=\rho_{kp}(t)dt \eea
 \noindent As it
is the case for the BGM model, we can use  a piecewise parametric
form or  a functional form for the serial volatilities
$\sigma_i(t)$ and the correlation $\rho_{ij}(t)$ (here full rank)
as \bea \sigma_i(t)&=&N_i [(a(T_{i-1}-t)+d)e^{-b(T_{i-1}-t)}+c] \;
\forall t \leq T_{i-1} \\
 \rho_{ij}(t=0)&=&\rho_L+(1-\rho_L)e^{-(\delta_A-\delta_ B \;
min[T_{i-1},T_{j-1}])|T_{i-1}-T_{j-1}|} \eea

\begin{table}[tpb]
\begin{center}
\begin{tabular}{|c|c|}
\hline BGM parameters & $a$, $b$, $c$, $d$, $\phi_i$, $\rho_L$,
$\delta_A$, $\delta_B$\\ \hline
Cev parameters & $\beta_i, \; i=1, \cdots, n$ \\
\hline SABR parameters &  $\alpha$, $\nu$, $\rho_{ia} \; i=1, \cdots, n$, \\
\hline
\end{tabular}
\caption{SABR-LMM: $9+3n$ parameters}
\end{center}
\end{table}

\noindent The constants $N_i$ are fixed such as $\sigma_i(0)=0$.
The model depends on $9+3n$ parameters (see Tab. 2) which are
calibrated on the swaption matrix. In the next subsection, we
derive the metric, the geodesic distance and the Abelian
connection underlying this model.

\subsection{Hyperbolic geometry}

\noindent By definition, the infinitesimal distance (at $t=0$)
between the point $x^\alpha$ and $x^\alpha+dx^\alpha$
(\ref{metric}) ($ds^2=\sum_{\alpha ,\beta=1}^{n+1} g_{\alpha
\beta} dx^\alpha dx^\beta$) is given by ($\rho^{ij}\equiv
[\rho^{-1}]_{ij}$ , $(i,j)=(1, \cdots ,n)$ and
$\rho^{ia}\equiv[\rho^{-1}]_{ia}$ are the components of the
inverse of the correlation matrix $\rho$) \bea ds^2={2 \over \nu^2
a^2}  ( \sum_{i,j=1}^n  \rho^{ij} {\nu dF_i \over C_i(F_i)} {\nu
dF_j \over C_j(F_j)} + 2 \sum_{i=1}^n \rho^{ia} {\nu dF_i \over
C_i(F_i)} da +\rho^{aa} da^2) \eea \noindent After some algebraic
manipulations, we show that in the new coordinates $[x_k]_{k=1
\cdots n+1}$ ($L$ is the Cholesky decomposition of the (reduced)
correlation matrix: $[\rho]_{i,j=1 \cdots n}=[ \hat{L}
\hat{L}^\dag]_{i,j=1 \cdots n}$)
 \bea x_k&=&\sum_{i=1}^n \nu \hat{L}^{ki}\int_{F_i^0}^{F_i} {dF'_i \over C_i(F'_i)}+\sum_{i=1}^n \rho^{ia} \hat{L}_{ik}
 a \; , \;  k=1,\cdots, n
 \\
x_{n+1}&=&(1-\sum_{i,j}^n \rho^{ia} \rho^{ja} \rho_{ij})^{1 \over
2} a \eea
\noindent the metric becomes
 \bea ds^2=
{2(1-\sum_{i,j}^n \rho^{ia} \rho^{ja}  \rho_{ij}) \over \nu^2  }
{\sum_{i=1}^n dx_i^2 +dx_{n+1}^2 \over x_{n+1}^2} \eea

\noindent  Written in the coordinates $[x_i]$, the metric is
therefore the standard hyperbolic metric on $\HH^{n+1}$ (modulo a
 constant factor ${2(1-\sum_{i,j,k=1}^n \rho^{ia} \rho^{ja} \rho_{ij} )  \over \nu^2  }$). In order to compute our
saddle-point (\ref{local1}), we need the geodesic distance which
is given by \cite{terras}

\begin{prop}
The geodesic distance $d(x,x')$ on $\HH^{n+1}$ is given by
 \bea
d(x,x^0)=cosh^{-1}[1+{\sum_{i=1}^{n+1} (x_i-x^0_i)^2 \over 2
x_{n+1} x^0_{n+1}}] \eea

\end{prop}

\noindent Using the geodesic distance on $\HH^{n+1}$ between the
points $x=(\{F\}_k,a)$ and the initial point
$x^0=(\{F^0\}_k,\alpha)$ ($ q_i=\int_{F_i^0}^{F_i} {dF'_i \over
C_i(F'_i)}$) given by
 \beaa d(x,x^0)={\sqrt{2}(1-\sum_{i,j}^n \rho^{ia} \rho^{ja}  \rho_{ij})^{1 \over 2} \over
\nu  }cosh^{-1}[1+{\nu^2 \sum_{i,j=1}^n \rho^{ij} q_i q_j +2 \nu
\sum_{j=1}^n \rho^{ja} q_j +(a-\alpha)^2 \over 2(1-\sum_{i,j=1}^n
\rho^{ia} \rho^{ja} \rho_{ij} )a\alpha}] \eeaa \noindent we derive
the following non-linear equations (\ref{local1}) satisfied by the
saddle-point $a^*(s),q_i^*(s)$ which implicitly depends  on $s$,
the swaption strike:
\begin{subequations} \label{sp}
\begin{gather}
a^*(s)=\sqrt{\alpha^2+2\nu \sum_{i=1}^n  \rho^{ia} q^*_i
+\nu^2 \sum_{i,j=1}^n \rho^{ij} q^*_i q^*_j } \label{sp1} \\
(2\nu \rho^{ia} +2\sum_{j=1}^n \nu^2 \rho^{ij} q^*_j)
{d(a^*,\{q^*_i\}) \over a^*(s)(cosh(d(a^*,\{q^*_i\}))^2-1)^{1
\over 2} }=-\lambda \alpha (1-\sum_{p,q=1}^n \rho^{pa} \rho^{qa}
\rho_{pq})  {\partial s_{\alpha \beta} \over
\partial q_i}|^* \label{sp2}
\end{gather}
\end{subequations}
\noindent with \bea cosh(d(a^*,\{q^*_i\})=1+{a^*(s)-\alpha \over
(1-\sum_{i,j=1}^n \rho^{ia} \rho^{ja} \rho_{ij} )\alpha}
\label{dd} \eea

\noindent and with \bea q^*_i&=&\phi_i^{-1} ({{F_i^*}^{1-\beta_i}
-{F_i^0}^{1-\beta_i} \over (1-\beta_i)})
 \; , \; \beta_i \neq 1 \label{saddle} \\
 q^*_i&=&\phi_i^{-1} ln({F_i^* \over F_i^0}) \; , \; \beta_i=1\eea

\noindent The saddle-point is determined  by solving these
non-linear equations (\ref{sp}) and an approximation (which could
be used as a guess solution in a numerical optimization routine)
is found by linearizing these equations around the spot Libor
rates (i.e. $q_i=0$)
\begin{subequations} \label{ap}
\begin{gather}
\lambda^*(s) \alpha^2 ={-2 \nu^2 (s-s_0) -2\nu \sum_{j=1}^n
\rho_{ij} \rho^{ja} \omega_j \over (1-\sum_{p,q=1}^n \rho^{pa}
\rho^{qa} \rho_{pq})\sum_{i,j=1}^n \rho_{ij} \omega_i
\omega_j } +o((s-s_0)^2) \label{ap1} \\
q_i^*(s)={\sum_{j=1}^n \rho_{ij}  \omega_j (s-s_0) \over
\sum_{p,q=1}^n \omega_p \omega_q \rho_{pq} } +o((s-s_0)^2)
\label{ap2}
\end{gather}
\end{subequations}

\noindent with $\omega_i \equiv {\partial s_{\alpha \beta} \over
\partial q_i}(q_i=0) $. Note that when the strike is close to
at-the-money, the saddle-points are close to the spot Libors and
$a^*=\alpha$.

\noindent Moreover, by using the explicit expression for the
hyperbolic distance, the Van-Vleck-Morette determinant is \bea
\Delta(F,a,\alpha)={d(a,F|\alpha) \over
\sqrt{cosh^2(d(a,F|\alpha))-1}}\eea

\subsection{Connection}

The Abelian connection is  given by (\ref{aaa}) \footnote{ \bea
{\mathcal A}^{i}&=&{a^2 \over 2}(b^i-{1 \over 2} C_i
\partial_i C_i)
\\
{\mathcal A}^a&=&-{\nu^2 a^2 b^a(F,t) \over 2} \eea}

\bea {\mathcal A}_i&=&{1 \over  C_i(F_i)}[ \sum_{j=1}^n \rho^{ij}(
{b^j(F,t) \over C_j(F_j)}-{ \partial_j C_j(F_j)
\over 2})-\nu \rho^{ia}   b^a(F,t)] \\
{\mathcal A}_a&=&{1 \over \nu } [\sum_{j=1}^n \rho^{aj}( {b^j(F,t)
\over C_j(F_j)}-{ \partial_j C_j(F_j) \over 2 })-\nu \rho^{aa}
b^a(F,t)] \eea

\noindent where we have used that \bea \sqrt{g}={2^{n+1 \over 2}
det[\rho]^{-{1 \over 2}} \over {\nu} a^2 \prod_{i=1}^n C_i(F_i)}
\eea

\noindent Finally, the Abelian 1-form connection is
 \beaa {\cal A}={1 \over \nu}\sum_ {j=1}^n[ ( {b^j(F,t) \over  C_j(F_j)}-{
\partial_j C_j(F_j) \over 2})(\nu \sum_{i=1}^n \rho^{ij} dq_i
+\rho^{aj}da)] -b^a(F,t)(\nu \sum_{i=1}^n \rho^{ia} dq_i
+\rho^{aa} da) \eeaa

\noindent In order to compute the log of the parallel gauge
transport $ln({\cal P})(a,q|\alpha)=\int_{\cal C} {\cal A}$, we
need to know a parametrization of the geodesic curve on
$\HH^{n+1}$. However, we can directly find $ln({\cal
P})(a,q|\alpha)$ if we approximate the drifts $b^k(F,t)$ by their
values at the Libor spots (and $t=0$). A similar approximation was
done in the Hagan-al formula \cite{sab} as was shown in
\cite{phl}. Modulo this approximation, \beaa ln({\cal
P})(a,q|\alpha) \sim {1 \over \nu }\sum_ {j=1}^n[ ( {b^j(F^0,0)
\over  C_j(F_j^0)}-{
\partial_j C_j(F_j^0) \over 2})(\nu \sum_{i=1}^n \rho^{ij} q_i
+\rho^{aj}(a-\alpha))]-b^a(F^0,0)(\nu \sum_{i=1}^n \rho^{ia} q_i
+\rho^{aa} (a-\alpha))  \eeaa

\subsection{Asymptotic Smile-Summary}
\noindent The asymptotic local volatility is given by
(\ref{localvol})

\beaa (\sigma_{loc}^{\alpha
\beta})^2(s,\tau)&=&\sum_{i,j=1}^n\rho_{ij} \sigma_i(t)
\sigma_j(t) f_{ij}(a,F)  (1+2\tau \sum_{\mu,\nu=1}^{n+1}A^{\mu
\nu} ( ({\partial_{\mu \nu} f_{ij}(a,F) \over f_{ij}(a,F)} +
2{\partial_{\mu} f_{ij}(a,F) \over f_{ij}(a,F)}{\partial_{\nu}
\psi(a,F) \over \psi(a,F)}) -\\&&\sum_{\gamma,\delta=1}^{n+1}
A^{\gamma \delta} {\partial_{\mu} f_{ij}(a,F) \over f_{ij}(a,F)}
\partial_{\nu \gamma \delta} d^2(a,F) ))|_{(a,F)=(a^*(s),F^*(s))}
 \eeaa

\noindent \noindent with $({a^*}^2(s),\{F^*_i\}_i(s))$ the
saddle-point satisfying the equations (\ref{sp}) and approximated
by (\ref{ap}) and

\beaa f_{ij}(a,F)&=&a^2 C_i(F_i) C_j(F_j) {\partial s_{\alpha
\beta} \over \partial F_i} {\partial s_{\alpha \beta} \over
\partial F_j} \; , \; \psi(a,F)=\sqrt{g \Delta}{\cal P} \; , \;
A^{\alpha \beta}=[\partial_{\alpha \beta} d^2]^{-1} \\
d(a,F)&=&{\sqrt{2}(1-\sum_{i,j}^n \rho^{ia} \rho^{ja}
\rho_{ij})^{1 \over 2} \over \nu }cosh^{-1}[1+{\nu^2
\sum_{i,j=1}^n \rho^{ij} q_i q_j +2 \nu \sum_{j=1}^n \rho^{ja} q_j
+(a-\alpha)^2 \over 2(1-\sum_{i,j=1}^n \rho^{ia}
\rho^{ja} \rho_{ij} )a\alpha}] \\
ln({\cal P})(a,q|\alpha) &\sim& {1 \over \nu }\sum_ {j=1}^n[ (
{b^j(F^0,0) \over  C_j(F_j^0)}-{
\partial_j C_j(F_j^0) \over 2})(\nu \sum_{i=1}^n \rho^{ij} q_i
+\rho^{aj}(a-\alpha))]-b^a(F^0,0)(\nu \sum_{i=1}^n \rho^{ia} q_i
+\rho^{aa} (a-\alpha)) \\
\Delta(F,a,\alpha)&=&{d(a,F|\alpha) \over
\sqrt{cosh^2(d(a,F|\alpha))-1}}\\
\sqrt{g}&=&{2^{n+1 \over 2} det[\rho]^{-{1 \over 2}} \over {\nu}
a^2 \prod_{i=1}^n C_i(F_i)} \eeaa

\noindent Note that this expression is {\it exact} when $\tau$
goes to zero. The smile at the first-order is then obtained by
plugging the above expression into (\ref{smile}).

\begin{rem}[Libor CEV model]
Note that our model for $\nu$ goes to zero (and $\alpha \equiv 1$)
gives the Andersen-Andreasen CEV libor model (with different CEV
parameters for each libors) and the above expressions degenerates
into \bea f_{ij}(F)&=& C_i(F_i) C_j(F_j) {\partial s_{\alpha
\beta} \over
\partial F_i} {\partial s_{\alpha \beta} \over
\partial F_j}\\ d(F)&=&\sqrt{2 \sum_{i,j=1}^n \rho^{ij}q_iq_j}  \\
ln({\cal P})(q)& =& \sum_ {j=1}^n ( {b^j(F^0,0) \over
C_j(F_j^0)}-{\partial_j C_j(F_j^0) \over 2})\sum_{i=1}^n \rho^{ij} q_i \\
\Delta(F,F^0)&=&1 \\
\sqrt{g}&=&{2^{n \over 2} det[\rho]^{-{1 \over 2}} \over
\prod_{i=1}^n C_i(F_i)}  \eea

\noindent with the saddle-points (\ref{sp}) satisfying the
non-linear equations (modulo the constraint $s_{\alpha \beta}=s$)
 \bea
\rho^{ij} q^*_j=-{\lambda \over 4}{\partial s_{\alpha \beta} \over
\partial q_i}|^*
\eea
\end{rem}

\subsection{Comments and Numerical Tests}

 \noindent It is
interesting to note that for $n=1$, i.e. for a caplet, the caplet
asymptotic smile reduces to the classical SABR formula by
construction.

\noindent Moreover, the asymptotic local volatility is given at
the zero-order by \beaa (\sigma^{\alpha \beta}_{loc})^2(s,t)
=\sum_{i,j=1}^n {\rho_{ij}(t) \sigma_i(t) \sigma_j(t)}
 {a^*}^2(F^*) C_i(F_i^*)
 C_i(F_i^*){\partial s_{\alpha \beta} \over \partial F_i}(F^*) {\partial s_{\alpha \beta} \over \partial
 F_j}(F^*)  \eeaa with $F^*$ depending implicitly on $s$ via
 (\ref{sp}). At this stage, it is useful to recall how a similar asymptotic
local volatility is derived using the "freezing" argument. The
forward swap rate satisfies the following SDE in the forward swap
num\'eraire $\QQ^{\alpha \beta}$ \bea ds_{\alpha
\beta}&=&\sum_{k=1}^n {\partial  s_{\alpha \beta} \over
\partial F_k} \sigma_k(t) a C_k(F_k) dZ_k \label{eq1}
\eea \noindent The "freezing" argument consists in assuming that
the terms ${\partial  s_{\alpha \beta} \over
\partial F_k}$ and ${C(s) \over C(F_i)}$ are almost constant.
Therefore, the SDE (\ref{eq1}) can be approximated by \bea
ds_{\alpha \beta}=\sum_{k=1}^n {\partial  s_{\alpha \beta} \over
\partial F_k}(F^0) \sigma_k(t) a {C_k(F_k^0) \over C_k(s^0)} C_k(s) dZ_k \eea
\noindent and the local volatility is \beaa (\sigma^{\alpha
\beta}_{loc})^2(s,t)=\sum_{i,j=1}^n \rho_{ij}(t) \sigma_i(t)
\sigma_j(t) {a^*}^2(s) {C_i(F_i^0) \over C_i(s^0)}
 {C_j(F_j^0) \over C_i(s^0)} {\partial s_{\alpha \beta} \over \partial F_i}(F^0)
 {\partial s_{\alpha \beta} \over \partial F_j}(F^0) C_i(s) C_j(s) \eeaa

\noindent  We can reproduce this formula for the swaption smile
at-the-money \footnote{An at-the-money swaption (ATM)  has a
strike $K$ equal to the spot rate $s_{\alpha \beta}(0)$ and an
out-of-the money (OTM) (resp. in-the-money (ITM)) swaption has
$K<s_{\alpha \beta}(0)$ (resp. $K>s_{\alpha \beta}(0)$).} as the
saddle-point Libor rates coincides with the spot rates. This is
not the case for in/out-the-money swaption. Therefore our
expression (exact at the zero-order) shows that the freezing
argument is no longer correct when we try to fit a swaption
implied smile in/out-the-money. In the following, we have tested
our asymptotic swaption formula at the zero-order (Formula F1)
against the Andersen-Andreasen asymptotic formula (Formula F2)
\cite{and} in the case $\nu=0$. The accuracy of these
approximations are examined using Monte-Carlo (MC) prices as a
benchmark. Following \cite{kaw}, we consider  five scenarii (see
Tables 3-4-5-6-7). In the following tables, the implied volatility
is reported and the numbers in brackets are the errors (in basis
points i.e. true volatility times $10^4$) corresponding to  the
implied volatility computed using the F1 or F2 formula minus the
MC implied volatility. An $x\times y$ swaption has an option
maturity of $x$ years, a swap length of $y$ years and a tenor of
one year. We set a time-step for Monte-Carlo $\delta=0.125$ and
$2^{16}$ paths \footnote{We have used a predictor-corrector scheme
with a Brownian bridge.}. Our formula F1 is more accurate than F2.

\begin{table}[h]
\begin{center}
\begin{tabular}{|c|c|c|c|c|}
\hline Swaption & strike & MC & F1 & F2 \\

\hline & 4\%(ITM) & 22.42\%  & 22.41\% (-1) & 22.61\% (19)

\\ \hline $5 \times 15$& 6\%(ATM)  & 20.33\% & 20.41\% (8) & 20.46\% (13)

\\\hline & 8\%(OTM) & 18.92\% & 18.93\% (1) &19.01\% (10)

\\ \hline \hline
& 4\%(ITM)  & 22.41\% & 22.51\% (11) & 22.67\% (26)  \\ \hline

$10 \times 10$ & 6\%(ATM) & 20.38\% & 20.41\% (3) & 20.50\% (12)  \\
\hline

& 8\%(OTM) & 18.93\% & 18.93\% (-1) & 19.05\%  (12) \\ \hline

\end{tabular}

\caption{Scenario A: Libor volatility
$\lambda_i(t)=5\%$. Libor $L_i(0)=6\%$. $\beta=0.5$  }
\end{center}
\end{table}

\begin{table}[h]
\begin{center}
\begin{tabular}{|c|c|c|c|c|}
\hline
  Swaption & Strike & MC & F1 & F2 \\ \hline
& 5.08\%(ITM) & 18.12\% & 18.20\% (8) & 18.17\%   (5)\\ \hline

$5 \times 15$ & 7.26\%(ATM) & 16.51\% & 16.61\% (10) & 16.63\%  (12) \\
\hline

& 9.44\%(OTM) & 15.38\% & 15.38\% (0) & 15.56\% (18)  \\ \hline
\hline

& 5.55\%(ITM) & 17.80\% & 17.81\% (1) & 17.89\%  (9) \\ \hline

$10 \times 10$ &7.93\%(ATM) & 16.26\% & 16.33\% (7) & 16.38\%   (11)\\
\hline

& 10.31\%(OTM) & 15.17\% & 15.19\% (2) & 15.32\% (15)  \\ \hline

\end{tabular}
\caption{Scenario B: Libor volatility
$dL_i=0.25(0.17+0.002(T_{i-1}-t))L_i^\beta dW$. Libor
$L_i(0)=log(a+bi)$, $L_0(0)=5\%$, $L_{19}(0)=9\%$ . $\beta=0.5$ }
\end{center}
\end{table}

\begin{table}[h]
\begin{center}
\begin{tabular}{|c|c|c|c|c|}
\hline
  Swaption & Strike & MC & F1 & F2 \\ \hline
& 5.08\%(ITM) & 14.89\% & 14.97\% (8) & 15.08\%   (19)\\ \hline

$5 \times 15$ & 7.26\%(ATM) & 13.73\% & 13.79\% (4) &13.81\%  (8) \\
\hline

& 9.44\%(OTM) & 12.92\% & 12.91\% (-1) & 12.92\% (0)  \\ \hline
\hline

& 5.55\%(ITM) & 14.52\% & 14.53\% (1) & 14.64\%  (12) \\ \hline

$10 \times 10$ & 7.93\%(ATM) & 13.33\% & 13.38\% (5) & 13.40\%   (7)\\
\hline

& 10.31\%(OTM) & 12.51\% & 12.51\% (0) & 12.54\% (3)  \\ \hline

\end{tabular}
\caption{Scenario C: Libor volatility
$dL_i=0.25(0.17-0.002(T_{i-1}-t))L_i^\beta dW$. Libor
$L_i(0)=log(a+bi)$. $L_0(0)=5\%$, $L_{19}(0)=9\%$. $\beta=0.5$ }
\end{center}
\end{table}

\begin{table}[h]
\begin{center}
\begin{tabular}{|c|c|c|c|c|}
\hline
  Swaption & Strike & MC & F1 & F2 \\ \hline

& 5.08\%(ITM) & 19.19\% & 19.33\% (14) & 19.38\%  (19) \\ \hline

$5 \times 15$ & 7.26\%(ATM) & 17.59\% & 17.72\% (13) & 17.75\% (16)  \\
\hline

& 9.44\%(OTM) & 16.46\% & 16.49\% (3) & 16.61\% (15)  \\ \hline
\hline

& 5.55\%(ITM) & 18.92\%  & 18.94\% (2) & 19.06\%  (14) \\ \hline

$10 \times 10$ & 7.93\%(ATM) & 17.31\%  & 17.39\% (8) & 17.45\% (14)  \\
\hline

& 10.31\%(OTM) & 16.18\% & 16.21\%(3)  & 16.32\% (14)  \\ \hline
\end{tabular}
\caption{Scenario D: $dL_i=0.05 L_i^\beta({b_{i1}(t) \over \sqrt{
b_{i1}(t)^2+b_{i2}(t)^2}} dW_1+{b_{i2}(t) \over \sqrt{
b_{i1}(t)^2+b_{i2}(t)^2}}dW_2)$. $b_{i1}(t)=\rho
e^{-k_1(T_{i-1}-t)}+\theta e^{-k_2(T_{i-1}-t)} $,
$b_{i2}(t)=\sqrt{1-\rho^2} e^{-k_1(T_{i-1}-t)}$. $\rho=0.99$,
$\theta=-0.99$, $k_1=k_2=0.54$. Libor $L_i(0)=log(a+bi)$.
$L_0(0)=5\%$, $L_{19}(0)=9\%$. $\beta=0.5$ }
\end{center}
\end{table}

\begin{table}[h]
\begin{center}
\begin{tabular}{|c|c|c|c|c|}
\hline
  Swaption & Strike & MC & F1 & F2 \\ \hline
& 5.08\%(ITM) & 33.09\% & 33.65\% (56)& 34.24\%(115)   \\ \hline

$5 \times 15$ & 7.26\%(ATM) & 29.47\% & 29.96\% (49) & 30.23\%(76)   \\
\hline

& 9.44\%(OTM) & 26.92\% & 27.14\% (22)& 27.49\%  (57) \\ \hline
\hline

& 5.55\%(ITM) & 31.75\% & 32.41\% (66)& 33.33\% (158)  \\ \hline

$10 \times 10$ & 7.93\%(ITM) & 28.47\% & 28.88\% (41)& 29.37\%(91)   \\
\hline

& 10.31\%(OTM) & 26.01\% & 26.18\% (17)& 26.68\% (67)  \\ \hline

\end{tabular}
\caption{Scenario E: Scenario D with
 $\beta=0.3$ }
\end{center}
\end{table}

\noindent In \cite{phl}, we explained how to derive a general
asymptotic smile for any stochastic volatility model using this
geometric framework. As an application, we derived an asymptotic
smile for a SABR model with a mean-reversion term. In the
following, we try to consider some natural extensions of our
SABR-BGM model where we add a non trivial drift to the volatility
process. The only modification comes from the expression of the
Abelian connection.

\subsection{Extensions}

Under the spot Libor measure, we assumed that the volatility
follows the  process  \bea da=-\nu a^2 \psi^a(a)dt+\nu a dZ_{n+1}
\eea \noindent with  $\psi^a(a)$ a general analytical function of
$a$ (the scaling $a^2$ in front  of $\psi^a(a)$ has been put for
convenience). \noindent After some algebraic computations, we
derive the new Abelian 1-form connection
 \beaa {\cal A}={1 \over \nu
}\sum_ {j=1}^n[ ( {b^j(F,t) \over  C_j(F_j)}-{
\partial_j C_j(F_j) \over 2})(\nu \sum_{i=1}^n \rho^{ij} dq_i
+\rho^{aj}da)]-(b^a(F,t) + \psi(a))(\nu \sum_{i=1}^n \rho^{ia}
dq_i+{\rho}^{aa} da) \eeaa

\noindent Using a similar approximation as before, i.e.
$C_j(F_j)\sim C_j(F_j^0)$ and $\psi^a(a) \sim \psi^a(\alpha)$, we
obtain for the parallel gauge transport \bea ln({\cal
P})(a,q|\alpha) \sim&& {1 \over \nu }\sum_ {j=1}^n[ ( {b^j(F^0,0)
\over C_j(F_j^0)}-{
\partial_j C_j(F_j^0) \over 2})(\nu \sum_{i=1}^n \rho^{ij} q_i
+\rho^{aj}(a-\alpha))]-(b^a(F^0,0) \nonumber\\&+&
\psi(\alpha))(\nu \sum_{i=1}^n \rho^{ia} q_i+{\rho}^{aa}
(a-\alpha)) \eea

\noindent Finally, the smile is obtained using our general formula
(\ref{smile}). Note that the metric and the geodesic equations
remain unchanged when we only modify the drift terms.

\section{Conclusion}
In this short note, we have introduced a LMM model coupled to a
SABR stochastic volatility process. By using the heat kernel
expansion technique in the short time limit, we have obtained an
asymptotic swaption implied volatility at the first-order,
compatible with the Hagan-al classical formula for caplets.
Moreover, we have seen that this exact expression (when the expiry
is very short) is incompatible with the analog expression obtained
using the freezing argument.

\end{document}